\documentclass[]{interact}

\usepackage{epstopdf}
\usepackage[caption=false]{subfig}

\usepackage[numbers,sort&compress]{natbib}
\bibpunct[, ]{[}{]}{,}{n}{,}{,}

\usepackage{colortbl}
\usepackage{xcolor,multicol}
\usepackage{pifont,bm,amsthm,mathrsfs}
\usepackage[nointegrals]{wasysym}
\usepackage{algorithmic,algorithm}
\usepackage{mathtools}
\usepackage{comment,amsmath}
\usepackage{amssymb}
\usepackage{float}
\newcommand{\E}{\mathbb{E}}

\newcommand{\Pb}{\mathbb{P}}

\usepackage{lmodern}

\newtheorem{remark}{Remark}[section]

\theoremstyle{plain}

\theoremstyle{definition}

\title{Parameter estimation and model selection for stochastic differential equations for biological growth}

\author{
\name{F. Baltazar-Larios\textsuperscript{a}\thanks{CONTACT F. Baltazar-Larios. Email: fernandobaltazar@ciencias.unam.mx}, F.J. Delgado-Vences\textsuperscript{b} and A. Ornelas Vargas\textsuperscript{c}}
\affil{\textsuperscript{a}Facultad de Ciencias, Universidad Nacional Aut\'onoma de M\'exico, M\'exico; 
\textsuperscript{b}Conacyt  Research Fellow, Instituto de Matem\'aticas, Universidad Nacional Aut\'onoma de M\'exico, Oaxaca,M\'exico; \textsuperscript{c}Conacyt  Research Fellow, Centro Interdisciplinario de Ciencias Marinas, Instituto Politecnico Nacional, La Paz, M\'exico.}
}
\begin{document}
\maketitle

\begin{abstract}
In this paper, we consider stochastic versions of three classical growth models given by ordinary differential equations (ODEs). Indeed we use stochastic versions of Von Bertalanffy, Gompertz, and Logistic differential equations as models. We assume that each stochastic differential equation (SDE) has some crucial parameters in the drift to be estimated and we use the Maximum Likelihood Estimator (MLE) to estimate them. For estimating the diffusion parameter, we use the MLE for two cases and the quadratic variation of the data for one of the SDEs. We apply the Akaike information criterion (AIC) to choose the best model for the simulated data. We consider that the AIC is a function of
the drift parameter. We present a simulation study to validate our selection method.

The proposed methodology could be applied to datasets with discrete observations including highly sparse data. Indeed, we can use this method even in the extreme case where we have observed only one point for each
path, under the condition that we observed a sufficient number of trajectories. For the last two cases, the data can be viewed as incomplete observations of a model with a tractable likelihood function; then, we propose a version of the Expectation Maximization (EM) algorithm to estimate these parameters. This type of dataset
typically appears in fishery, for instance.
\end{abstract}


\section{Introduction}\label{sec1}

 The main motivation for this work comes from problems where it is necessary to fit SDEs to modelling growth biological data; for instance in marine biology, ecology (see \cite{de-Gross}), oncology (see \cite{Paek-14}), or in paleontology (to model sclerochronological parameters of shell growth \cite{moss-etal}). Given that real systems cannot be completely isolated from their environments and, therefore, always experience external stochastic forces, the use of stochastic differential equations (SDEs) as models is justified and preferred to ordinary differential equations (ODEs). SDEs have been used in several fields such as finance, econometrics, population systems, ecology, etc.; thus, fitting SDEs to actual data has a wide interest. 
 
Fitting differential equations is a classic example in inverse problems and has been treated by many authors with several approaches (see for instance \citep{ha-ro}, \cite{isakov}, or \cite{Lilla-10}) in particular, fitting ODEs have been deeply studied. One typical situation in these problems is when fitting ODEs is equivalent to adjusting only a fixed number of unknown and constant parameters. Motivated by this idea, we will assume that we have this case, meaning we assume that every SDE has some crucial parameters that we focus on estimating (see  \cite{Chao-16} or \cite{miao}).
 
In this work, we consider the problem of estimating the parameters using the available data to fit each SDEs and propose a criterion to choose the best fit. There exists theory to do this estimation and, depending on the observed data available,  could be applied to solve the problem. However, depending on the available data, it is possible to use one of the existing methods to estimate parameters (see, for instance, \cite{iacus}, \cite{panik}). 

Based on the available data, the three scenarios we consider in this work are:
\begin{itemize}
    \item  a complete (continuous) observation of the solutions,
    \item discrete observations (highly sparse data), and 
    \item only one observed measurement for a sufficiently large number of individuals.
\end{itemize}

Continuous observation is unrealistic because very few phenomena can be continuously observed or recorded, then we consider continuous observation when the discrete observations are high frequency. The second case is the most common one, although it is known that insufficient data can lead to extremely inaccurate parameter estimation. In the last scenario, we assume that the data is a collection of several individuals coming from the same population and under the same environmental conditions, which allows us to mix the observation to simulate discrete observations. These assumptions allow us to estimate the parameters with good precision. 
This type of data appears usually in the fishery.

In this work, the last two scenarios are considered as incomplete observations of the first scenario and then we propose the use the EM algorithm for complete data.

In summary, in this work, we fit some examples of SDEs into the three scenarios of observations described above. In addition, the three SDEs considered in this research are stochastic versions of classical ODEs, which are widely employed for modelling biological or population growth phenomena. These SDEs include the logistic, Von Bertalanffy, and Gompertz equations (cf. \cite{panik}). Nevertheless, the method used in this paper could be applied to other SDEs.

A statistical model is a useful tool for the description and forecast of a stochastic (or deterministic) system. We could roughly say that a statistical model is a simplification of
a complex reality. We could think that the complex reality is a “true” model and one of the goals of mathematicians and statisticians is to approximate (in some sense) this “true”  model.  Depending on the problem, we might assume that the “true” model is contained within a set of models under our consideration \cite{ando-10}. To choose the best model assuming a given dataset could be done using several methods, the most used is Akaike’s information criterion, known as AIC, for evaluating the constructed models. AIC has been applied successfully to linear regression, time series, etc.

The stochastic models we focus on in this paper are Gompertz, Von Bertalanffy, and Logistic stochastic differential equations. We consider that some parameters in each SDE are constants but unknown, usually, the drift parameter \footnote{the function space-time of deterministic part of the model.} will represent the intrinsic growth rate of the model; thus, we are interested in estimating this parameter. Given the data, we are interested in estimating the parameters for each SDE and we apply the AIC to each of the three SDEs which provides a criterion to choose the best model that fits to data.

Parameter estimation for SDEs using the MLE method (continuous observation), the EM algorithm, the Ozaki method, and Bayesian methods (discrete observation), have been studied in \cite{panik} and \cite{iacus}, for instance. In this work, we calculate the likelihood function of the transition probabilities for the parameter of the Gompertz and Von Bertalanffy models obtain the MLEs for the parameters in the continuous case, and use the EM algorithm to obtain the corresponding MLEs for the case of incomplete information (discrete observation). For the Logistic model,  we follow the method for estimating the parameters described in \cite{de-etal-22}, where the estimation combines quadratic variation of the observations and MLE via the EM algorithm. For the case where there is only a dataset corresponding to one record for each individual, but with observations of several different individuals  at different points in time, we propose a novel algorithm to simulate paths of the SDEs and obtain the corresponding parameters.

Since we know that by maximizing the log-likelihood, it is possible to obtain the estimators and we conclude that we have good estimators for Gompertz and Von Bertalanffy models. For the Logistic model, the properties of the estimators were proved in \cite{de-etal-22}.

It is known that the AIC is a function of the log-likelihood, which depends on the parameters of the model. In this paper, we use the MLEs (Gompertz and Von Bertalanffy) to calculate the AIC. For the Logistic model, we assume that the log-likelihood is a function only of the drift parameters, meaning that the diffusion parameter is previously fixed, afterwards we calculate the corresponding AIC. We validate this method using simulated data for the three SDEs and, then we fit the best stochastic model.

 This paper is organized as follows. In Section \ref{sec-3models} we present the SDEs considered in this manuscript, Section \ref{sec-MLE}  contains the parameter estimation under three types of datasets. Section \ref{sec-simulation} presents the simulations study for the different scenarios and the numerical result of this simulation. In Section \ref{sect_model_sel} we apply the AIC criteria to the SDEs and we present the results of numerical simulations. Section \ref{sec-Conclusions} includes the conclusions and final remarks of the work.

\section{Three stochastic growth models}\label{sec-3models}
We consider the one-dimensional, time-homogeneous SDE 
\begin{align}\label{SDE}
 dX_t &= \alpha_{\theta}(X_t) dt + \beta_{\theta}(X_t)dW_t,
\end{align}
where $W_t$ is a standard Wiener process, $\theta$ is the multidimensional parameter to be estimated, and $\alpha$ and $\beta$ are known functions such that the solution of \eqref{SDE} exists. We will consider three particular functions for $\alpha$ and $\beta$ and, therefore,  we will obtain three stochastic growth models, which are the subject of this manuscript.

\subsection{A stochastic Gompertz model}\label{ssec-SGDE}

Consider the stochastic Gompertz differential equation given by

\begin{align}\label{SGE}
dX_t &= -b X_t \log(X_t) dt + \sigma X_t dW_t, \qquad t>0\\
X_0&=x_0,\nonumber
\end{align}
i.e., we have taken $\alpha_{\theta}(X_t)=-bX_t \log(X_t) $, $\beta_{\theta}(X_t)=\sigma X_t$, and $\theta=(b,\sigma)$, where $b$ and $\sigma>0$ are constants. In this equation, we are setting the carrying capacity, or in our case the maximum size that can be reached by the spices, equal to 1.

The solution to \eqref{SGE} is obtained by applying the It\^o formula to $Y_t=\log(X_t)$. Indeed, we have that the stochastic process $Y_t$ solves
\begin{align*}
dY_t &= d\log(X_t)=\Big[-b \log(X_t)-\frac{\sigma^2}{2} \Big] dt + \sigma dW_t\\
&=  \Big[-\frac{\sigma^2}{2} -b Y_t  \Big] dt + \sigma dW_t,
\end{align*}
which we identify as the Ornstein-Uhlenbeck process for $Y_t=\log(X_t)$. This implies (see \cite{pavliotis} for further reading) that the solution is

$$
\log(X_t)=Y_t= Y_0e^{-bt} + \frac{-\sigma^2}{2b}\big(1-e^{-bt}\big) + \sigma \int_0^t e^{-b(t-s)}
dW_s,
$$
where $Y_0=\log(X_0)$.\\

Thus, the solution to \eqref{SGE} is

\begin{equation}\label{sol_SGE}
X_t= \exp\Bigg[\log(X_0)\, e^{-bt} - \frac{\sigma^2}{2b}\big(1-e^{-bt}\big) + \sigma \int_0^t e^{-b(t-s)} 
dW_s\Bigg].
\end{equation}
Recall that the stochastic integral is Gaussian and that if $Z\sim \mathcal{N}(m,\sigma^2)$ then  $\E\big[exp(tZ)]= e^{m t} e^{\sigma^2 t/2}$ . Using these facts, we can see that 
\begin{equation}\label{E_sol_SGE}
\E\big(X_t\big)= \exp\Bigg[\log(X_0)\, e^{-bt} - \frac{\sigma^2}{2b}\big(1-e^{-bt}\big)
+  \frac{\sigma^2}{4b}\big(1-e^{-2bt}\big)\Bigg],
\end{equation}

and then,

\begin{equation*}
\lim_{t\rightarrow\infty}
\E\big(X_t\big)= \exp\Big[-\frac{\sigma^2}{4b}\Big].
\end{equation*}

\begin{remark}
We observe that taking $\sigma=0$ in  \eqref{SGE} we recover a deterministic ordinary differential equation. Furthermore, using expression \eqref{E_sol_SGE} we can, heuristically, find a solution for the deterministic version of  \eqref{SGE} given by $\exp\Big[C \, e^{-bt}\Big]$, which agrees with the classical one.

\end{remark}
\subsection{A stochastic Von Bertalanffy model}\label{ssec-SVBDE}

We take $\alpha_{\theta}(L_t)=\kappa (L_{\infty}-L_t)$ and $\beta_{\theta}(L_t)=\sigma(L_{\infty}-L_t)$, thus we obtained the stochastic Von Bertalanffy differential equation, which is the SDE  given by

\begin{align}\label{S-VBE}
d L_t &= \kappa (L_\infty-L_t) dt + \sigma (L_\infty-L_t)dW_t, \qquad t>0,\\
L_0&=l_0,\nonumber
\end{align}
where we assume $L_\infty$to be  known,  $\theta=(\kappa,\sigma)$. Here  $\kappa$ is interpreted as the growth coefficient, which we assume is constant but unknown, and  $\sigma$ is the diffusion parameter.

By applying It\^o formula to the function $G_t= (L_\infty-L_t) $  we get 

\begin{align*}
dG_t &= d (L_\infty-L_t) = -\kappa (L_\infty-L_t) dt -\sigma (L_\infty - L_t) dW_t\\
&= -\kappa G_t - \sigma G_t dW_t.
\end{align*}

From the last equality, we identify $G$ as a Geometric Brownian motion with solution (see \cite{pavliotis})  given by 

\begin{align}\label{GB-sol}
G_{t}=G_{0}\exp \left(\left(-\kappa -{\frac {\sigma ^{2}}{2}}\right)t- \sigma W_{t}\right),
\end{align}

with $G_0=(L_\infty-l_0).$

From the explicit expression for $G_t$ we get that 

\begin{align}\label{sol_SVB}
L_t &=  L_\infty- (L_\infty-l_0) \exp \left(\left(- \kappa -{\frac {\sigma ^{2}}{2}}\right)t- \sigma W_{t}\right).
\end{align}

Note that 
$$
\lim_{t\rightarrow \infty}\E(L_t) = L_\infty.
$$

\subsection{A stochastic logistic model}\label{ssec-SLDE}

In this section, we present the SLDE which is obtained when we set $\alpha_{\theta}(P_t)= r P_t(1-P_t)$ and $\sigma_{\theta}(P_t)=\sigma P_t$, we have
\begin{equation}\label{sto-log}
  dP_t=  r P_t(1-P_t) dt + \sigma P_t dW_t,
\end{equation}
where   $\sigma>0$ and $p_0=P_0$ is a bounded absolutely continuous random variable $p_0(\omega):\Omega \rightarrow [a_1,a_2]\subset (0,1)$.  In the model \eqref{sto-log} allows the noise to depend on the size of the corresponding population. This equation has been studied in \cite{panik}. 

It is well-known that the deterministic version of the equation \eqref{sto-log} has been a good model for several phenomena (see for instance \cite{br-go} or \cite{br-et-al-83} and the references therein). In our case, $P_t$ will denote the proportional size of the individual of a given population, with $p_0$ being the random initial size. $r>0$ is the  intrinsic  growth rate of the species, and we assume it is our interest parameter to be estimated.


The strong solution (see \cite[see Th. 2.2. therein]{ji-shi})  to \eqref{sto-log} given by 
\begin{equation}\label{sol-sto-log}
 P_t=\frac{ f_t}{\frac{1}{p_0} +  r\int_0^t  f_s ds},
\end{equation}
where 
$$
f_t:=\exp\Big(t\big(r-\tfrac{\sigma^2}{2}\big)+ \sigma W_t \Big).
$$
We have that 
$$
\lim_{t\rightarrow\infty}  \E f_t= e^{rt},
$$
and, we deduce that 
$$
\lim_{t\rightarrow\infty}  P_t= 1.
$$

\section{Parametric estimation}\label{sec-MLE} 
In this section, we will assume that the parameters in each SDE are unknown and that the initial condition, for each SDE, is a random variable with some given density. In addition, we assume this density is the same for all possible values of the parameter. Fixed a positive time $T>0$ and a time interval $[0,T]$.

We present algorithms to estimate the parameters of equations \eqref{SGE},   \eqref{S-VBE}, and \eqref{sto-log} in three different scenarios.  First, we assume that we observe the data at continuous times in $[0,T]$, complete observation. The second case is when the observations are given at discrete time. Finally, we observed only one single point for each path for a suitable number of trajectories. The last case means that each observation represents a unique individual from the population in our model. 
 
 For the two last scenarios we use the EM algorithm to estimate the parameters (see  \cite{ce-di-86}, \cite{de-La-lr}, and \cite{ni-00}) by completing the information; that is, we see the gap between two consecutive points as missed information, and by using diffusion bridges we want to complete the information. Here, we are assuming that the models have a tractable likelihood function. In the EM algorithm, it is necessary to calculate the conditional expectation in the E-step, which is done using the approximation of diffusion bridges given in  \cite{bla-sor-14}. Thus, we calculate the MLE for the drift parameter in each SDE and we use the asymptotic properties of the MLE. 
 
 Notice that the noise in each SDE takes a multiplicative form; indeed, it is an affine function of the solution of the SDE multiplied by the Wiener process. Therefore, with the previous assumptions, the diffusion parameter can be estimated from the quadratic variation when the MLE corresponding cannot be determined. 
 
\subsection{Continuous case or complete data}\label{sec-MLE-cc}
For the case of complete observation in the interval time $[0,T]$, we assume that the observation interval is divided into $n$ sub-intervals $[t_{i-1},t_i]$ of length $\Delta_n=\frac{T}{n}$ with $0=t_0<t_1<\ldots<t_n=T$. Thus, to denote that we observe the paths continuously, we assume that $n$ is large enough such that $\Delta_n$  goes to zero.

\subsubsection{MLE for a stochastic Gompertz model}\label{sect_MLE-SGDE}
To compute the MLE  of parameters $\theta=(b,\sigma)$ of the SDE \eqref{SGE}, under complete observation in $[0,T]$, we use the solution given in \eqref{sol_SGE} and we take advantage that the likelihood function has a closed form  given by 
\small{\begin{align}\label{lik_gom}
L(y_0,y_1,\ldots,y_n;b,\sigma)=\prod_{i=1}^n\left[\frac{\exp\Bigg({-\frac{(y_i-(y_{i-1}e^{-b\Delta_n}-(\sigma^2/2b)(1-e^{-b\Delta_n})))^2}{\sigma^(1-e^{-2b\Delta_n})/b}\Bigg)}}{\sqrt{2\pi\sigma^2(1-e^{-2b\Delta_n})/b}}\right], 
\end{align}}
where $Y_{t_0}=y_0,Y_{t_1}=y_1,\ldots,Y_{t_n}=y_n$.
The MLEs are obtained by finding the maximum of the function \eqref{lik_gom}. Thus, the MLEs are 
\begin{equation}\label{b}
\hat{b}=-\frac{c_1}{\Delta_n},
\end{equation}
and
\begin{equation}\label{sigma_gom}
\hat{\sigma}=\sqrt{\frac{2c_3h\hat{b}}{1-e^{-2\hat{b}}}}\, ,
\end{equation}
where
\begin{eqnarray}\label{cs}
c_1&=&\frac{n\sum_{i=1}^ny_iy_{i-1}-\sum_{i=1}^ny_i\sum_{i=1}^ny_{i-1}}{n\sum_{i=1}^ny_{i-1}^2-(\sum_{i=1}^ny_{i-1})^2},\nonumber\\
c_2&=&\frac{\sum_{i=1}^ny_{i-1}-c_1\sum_{i=1}^ny_i}{n},\nonumber    \\
c_3&=&\frac{\sum_{i=1}^n(y_i-c_1y_{i-1}+c_2)}{n}.
\end{eqnarray}

\subsubsection{MLE for a stochastic Von Bertalanffy model}\label{sect_MLE-SVBDE}
For this model, we have the observations $\{L_{t_0},L_{t_1},\ldots,L_{t_n}\}$. Assuming we know the value of $L_{\infty}$, then we can obtain the corresponding observations $G_{t_i}=L_{\infty}-L_{t_i}$. By \eqref{GB-sol}, the conditional density function of the $\{G_t\}_{t=0}^T$ is log-normal and then the transition probabilities are 
\begin{align*}
p_{\Delta_n}(g_i\shortmid g_{i-1})=\frac{\exp\Bigg({-\frac{[\log(g_i)-(\log(g_{i-1})+(\kappa-\sigma^2/2)\Delta_n)]^2}{2\sigma^2\Delta_n}\Bigg)}}{g_i\sigma\sqrt{2\pi\Delta_n}},
\end{align*}
where $G_{t_i}=g_i$, $i=0,1,\ldots,n$. We have that the likelihood function is

\begin{align}\label{lik_GM}
L(g_0,g_1,\ldots,g_n;\kappa,\sigma)=\prod_{i=1}^n\left[\frac{\exp\Bigg(-\frac{[\log(g_i)-(\log(g_{i-1})+(\kappa-\sigma^2/2)\Delta_n)]^2}{2\sigma^2\Delta_n}\Bigg)}{g_i\sigma\sqrt{2\pi\Delta_n}}\right]. 
\end{align}
From \eqref{lik_GM} we see that the maximum likelihood estimators are
\begin{align}\label{sigma_van}
\hat{\sigma}^2=\frac{-a^2+2ab-b^2+cn-2dn+en}{nT}. 
\end{align}
\begin{align}\label{kappa_van}
\hat{\kappa}=\frac{b-a}{T}-\frac{\hat{\sigma}^2}{2}. 
\end{align}
where
\begin{eqnarray}\label{cons_von}
  a&=&  \sum_{i=1}^n\log(g_i),    \nonumber\\
   b&=&  \sum_{i=1}^n\log(g_{i-1}),   \nonumber \\
  c&=&  \sum_{i=1}^n\log^2(g_i),   \nonumber \\
   d&=&  \sum_{i=1}^n\log(g_{i-1})\log(g_i), \nonumber\\
   e&=&  \sum_{i=1}^n\log^2(g_{i-1}).    
\end{eqnarray}

\subsubsection{A stochastic logistic model}\label{sect_MLE-SLDE}

For this model, we use the quadratic variations to obtain an estimator of $\sigma$. The MLE for $r$ is presented in \cite{de-etal-22}, where the authors proved that the MLE is strongly consistent and asymptotically normal. Let $P_{t_0}=p_0,P_{t_1}=p_1,\ldots,P_{t_n}=p_n$ be the observations in the interval time $[0,T]$. Following \cite{de-etal-22}, we have that an estimator of $\sigma$ is given by 
\begin{align}\label{sigma_con_log}
\hat{\sigma}^2=\frac{2\sum_{i=1}^n(p_i-p_{i-1})^2}{\sum_{i=1}^n\big( p_i^2-p_{i-1}^2\big)}.
\end{align}
If we estimate $\sigma$ using the expression \eqref{sigma_con_log} we have that the full log-likelihood function of $r$ is 
\begin{align*}
l(r) &=r\frac{1}{\sigma^2 } \int_0^T \frac{\big( 1-P_t \big)}{ P_t} dP_t-\frac{1}{2} r^2\frac{1}{\sigma^2 } \int_0^T  \big( 1-P_t^2 \big) dt.
\end{align*}
The MLE of $r$ using the data is 
\begin{align}\label{r_con_log}
\hat{r}=\frac{\int_0^T \frac{\big( 1-P_t \big)}{ P_t}dP_t}{\int_0^T  \big( 1-P_t^2 \big) dt}\approx\left(\frac{1}{\sum_{i=1}^n(1-p_i)^2\Delta_n}\right)
\left(\sum_{i=1}^n\frac{(1-p_i)(p_i-p_{i-1})}{p_i}\right).\end{align}

\subsection{Discrete time observations}\label{sec-Discrete}

In this case, to estimate the parameters, we suppose that the available data is the set of observations of a realisation of the process at times $0=s_0<s_1<\cdots<s_{n-1}<s_n=T$, where $\Delta_i=s_i-s_{i-1}$,  ($i=1,\ldots,n$) and  $n$ is small such that $\{X_{s_0},X_{s_1},\ldots,X_{s_n}\}$ does not guarantee the existence of reliable parameter estimators. Then, we consider that $\{X_{s_0},X_{s_1},\ldots,X_{s_n}\}$ are  discrete observations for a continuous process $X$. We can consider the available dataset as an incomplete observation of a complete dataset given by the full data given in the last scenario. Then, we use diffusion bridges to complete the information in step E of the EM algorithm and the likelihood function of discrete observations (Gompertz and Von Bertalanffy models) or quadratic variation and MLE (logistic model) in step M of the EM algorithm to find the corresponding estimators. To do so, we should calculate the conditional expectation of the corresponding likelihood function for the model given the observations. Then, we need to simulate paths of the diffusion process given the data, i. e., simulate diffusion bridges.

A diffusion bridge from state $a$ at time $t_1$ to state $b$  at time $t_2$  $((a,t_1,b,t_2)-bridge)$ is a solution $\{X_t\}_{t=t_1}^{t_2}$ of a SDE such that $X_{t_1}=a$ and $X_{t_2}=b$. 

To calculate the conditional expectation in E-step of the EM algorithm, we use the estimator $\theta_k$ from the previous step and we generate a diffusion bridge of size $L_i$ between each consecutive observations, i.e., 
\begin{align*}
(X_{t_{i-1}},t_{i-1},X_{t_i},t_i)\approx \{X_{t_{i-1}}=x_{t_{i0}},x_{t_{i1}}\ldots,x_{t_{iL_i-1}},X_{t_{i}}=x_{t_{iL_i}}\}, 
\end{align*}
{where $\Delta_{L_i}=t_{i(l-1)}- t_{il}=\frac{\Delta_i}{L_i}$ for $l=1,\ldots,L_i$ and $i=1,\ldots,n$. We choose each $L_i$ as that $\Delta_{L_i}=\Delta_{L_j}=\Delta$ for all $i,j=1,\ldots,n$ and with $\Delta$ close to zero.

\subsubsection{Gompertz model}

We use the expressions \eqref{b}  and \eqref{sigma_gom} in an EM algorithm to estimate MLEs corresponding. The EM algorithm works as follows. Let $\theta_0=(b_0,\sigma_0)$  be the initial values of the parameters.

\begin{algorithm}[H]
\begin{enumerate}
\item  Set $k=0$ and choose $\theta_0=(b_0,\sigma_0)$.
\item \textbf{E-step}. Calculate $\mathbb{E}_{\theta_k}[Y_t \mid Y_{t_{i-1}},Y_{t_i}]$ for $t\in[t_{i-1},t_i]$ for $i=1,\ldots,n$. 
\item  \textbf{M-step}. Using the diffusion brides $\{(Y_{t_{i-1}},t_{i-1},X_{t_i},t_i)\}_{i=1}^n$ of E-step to calculate the constants \eqref{cs} we make    
\begin{eqnarray*}
b_{k+1}&=&-\frac{c_1}{\Delta_n},\\
\sigma_{k+1}&=&\sqrt{\frac{2c_3h{b_{k+1}}}{1-e^{-2b_{k+1}}}}.
\end{eqnarray*}
 \item  k=k+1 and go to 2. 

\end{enumerate}
  \caption{EM for the Gompertz model}\label{EM-dis-gom}
\end{algorithm}

Algorithm \ref{EM-dis-gom} runs $K$ iterations with a suitable burn-in of $K_0<K$ and then the  estimators are given by
\begin{equation}\label{est-dis-time}
\hat{b}_{ML}=\frac{\sum_{k=K_0}^Kb_k}{K-K_0},\hspace{.5cm}\mbox{and}\hspace{.5cm}\hat{\sigma}_{ML}=\frac{\sum_{k=K_0}^K\sigma_k}{K-K_0}.
\end{equation}
\subsubsection{Von Bertalanffy model}
Following the idea of previous sections for the case when the observations of paths are recorded at discrete times, we use expressions \eqref{sigma_van}, \eqref{kappa_van} and the EM algorithm to calculate the MLEs. Let $\theta_0=(\kappa_0,\sigma_0)$  be the initial values of the parameters, the EM algorithm corresponding to this model is as follows.

\begin{algorithm}[H]
\begin{enumerate}
\item  Set $k=0$ and choose $\theta_0=(\kappa_0,\sigma_0)$.
\item \textbf{E-step}. Calculate $\mathbb{E}_{\theta_k}[G_t\mid G_{t_{i-1}},G_{t_i}]$ for $t\in[t_{i-1},t_i]$ for $i=1,\ldots,n$. 
\item  \textbf{M-step}. Using the diffusion brides $\{(G_{t_{i-1}},t_{i-1},G_{t_i},t_i)\}_{i=1}^n$ of E-step to calculate the constants \eqref{cons_von}  and update the estimators 
\begin{eqnarray*}
\kappa_{k+1}&=&\frac{b-a}{T}-\frac{\hat{\sigma}^2}{2},\\
\sigma_{k+1}&=&\sqrt{\frac{-a^2+2ab-b^2+cn-2dn+en}{nT}}.
\end{eqnarray*}
 \item  k=k+1 and go to 2. 

\end{enumerate}
  \caption{EM for the Von Bertalanffy model}\label{EM-dis-von}
\end{algorithm}

Algorithm \ref{EM-dis-von} runs $K$ iterations with a suitable burn-in of $K_0<K$ and then the  estimators are given by
\begin{equation}
\hat{\kappa}_{ML}=\frac{\sum_{k=K_0}^K\kappa_k}{K-K_0},\hspace{.5cm}\mbox{and}\hspace{.5cm}\hat{\sigma}_{ML}=\frac{\sum_{k=K_0}^K\sigma_k}{K-K_0}.
\end{equation}
\subsubsection{Logistic model}
We follow \cite{de-etal-22} to  estimate $\sigma$, and as in the previous models, $r$ is estimated using diffusion bridges to complete information. The expressions (\ref{sigma_con_log}) and \eqref{r_con_log} are used  in the following EM algorithm.

\begin{algorithm}[H]
\begin{enumerate}
\item  Set $k=0$ and choose $\theta_0=(r_0,\sigma_0)$.
\item \textbf{E-step}. Calculate $\mathbb{E}_{\theta_k}[P_t\mid P_{t_{i-1}},P_{t_i}]$ for $t\in[t_{i-1},t_i]$. 
\item  \textbf{M-step}.

    $$r_{k+1}=\frac{\sum_{i=1}^n{s_i}}{n}.$$
    \item  \textbf{Update $\sigma$}
$$\sigma_{k+1}=\frac{2\sum_{j=1}^n \sum_{l=1}^L \big[p_i(t_{lj})-p_i(t_{(l-1)j})\big]^2}{\sum_{j=1}^n \sum_{l=1}^L \big[p_i(t_{lj})^2+p_i(t_{(l-1)j})^2\big]\Delta_L}.$$

 \item  k=k+1 and go to 2. 

\end{enumerate}
  \caption{Estimation of $\theta$}\label{EM1}
\end{algorithm}
To calculate the conditional expectation in the E-step of Algorithm \ref{EM1} we use the current $\theta_k$ and we generate a diffusion bridge to calculate $s_i$'s
\begin{align*}
s_{i} &= \frac{1}{\sum_{j=1}^n \sum_{l=1}^L \big( 1-p_i(t_{l(j-1)})\big)^2  \Delta_L  }\\
&\qquad\qquad\qquad \times \sum_{j=1}^n \sum_{l=1}^L \frac{\big( 1-p_i(t_{l(j-1)}) \big)}{ p_i(t_{l(j-1)})} \big[ p_i(t_{lj})- p_i(t_{(l-1)j})\big].
\end{align*}
To update $\sigma$ in Step 4, we use the continuous paths generated by the diffusion bridges for the E-step. Steps 2-5 are repeated until the convergence.

\subsection{One record for each path}\label{one-observation}

In this scenario, we consider a dataset with only one observation for each individual at a different time, which is a common situation in fisheries, for example (see \cite{de-etal-22}). However, such a dataset can also arise in various other phenomena where obtaining multiple observations for each individual is extremely challenging (for instance in longitudinal healthcare data, cf. \cite{la-etal-22}). To accurately estimate the parameters in the SDE, we will assume  that each individual comes from the same population, indicating that they are subject to identical environmental conditions and then we can assume that each path comes from the same stochastic process. We do not assume that the simulated paths have the same initial condition. In fact, we assume that they follow a beta distribution\footnote{which can be fitted using the time reversal SDE and its estimated parameters}. Following the idea of methods for simulating paths of diffusion processes and by combining all the data, we can create simulated trajectories given an observation at a certain time, using that they are non-decreasing and the variance of observations at the next time. Therefore, we could apply the method presented in the previous subsection. 

Algorithm  \ref{OR} is a new version of the method proposed in  \cite{de-etal-22} to generate an observed path at times $\{t_0,t_1,\ldots,t_n\}$ when we have only one measurement from each of  a suitable number of $M$ paths of the solution of the same SDE. 
 
We use known $\theta$, the parameter of SDE \eqref{SDE}, and the constants  $\alpha,\beta>0$ to generate a path $\{X_{t_i}\}_{i=0}^n$ of the dataset $\{\bm{X}^1,\ldots,\bm{X}^M\}$ ($\bm{X^m}=\{X^m_{t_0},X^m_{t_1},\ldots,X^m_{t_n}\}, m=1,\ldots,M$). The algorithm works as follows.

\begin{algorithm}[H]
  \begin{enumerate}
  \item  Draw $X^m_{t_0}$ from a Beta$(\alpha,\beta)$ distribution for $m=1,2,\ldots,M$.  

 \item  Create a partition of the time interval $[0,T]$ into $n$ subintervals of length $\Delta_n$, here we choose $n$ large enough such that  $\Delta_n$ goes to zero.

\item Using the Milstein scheme in the partition of the last step, draw $M$ paths for a process that is a solution of Equation (\ref{SDE}) with parameter $\theta$  in the time interval $[0,T]$. Let $X^m_{t_i}$ be the $i$th point of the $m$th path for $i=0,1,\ldots,n$ and $m=1,2,\ldots,M$.

 \item  Make $k=1$ and randomly choose $X_{t_0}$ from $\{X^1_{t_0},\dots,X^M_{t_0}\}$

\item Calculate the variance $v_k$ of $\{X^1_{t_{k}},\dots,X^M_{t_{k}}\}$ and drawn $Y_k\sim TN_{(X_{t_{k-1}-v_k},\infty)}(X_{t_{k-1}},v_k)$.

 \item Make $X_{t_k}=\mbox{inf}_{i}\{X^i_{t_{k}}\shortmid X^i_{t_{k}}\geq Y_k\}$.

\item If $k=n$, stop and the path is $\{X_{t_0},X_{t_1},\dots,X_{t_n}\}$. Otherwise,  $k=k+1$ and go to Step 5.

\end{enumerate}
  \caption{Sample one record}\label{OR}
\end{algorithm}
In Step (5) of Algorithm \ref{OR}, $TN_{(a,b)}(\mu,\gamma)$ denotes a Truncated Normal random variable in $(a,b)$ with mean $\mu$ and variance $\gamma$.
In this case, we use Algorithm \ref{OR} to generate paths and we can use methods proposed in the sections \ref{sec-MLE-cc} and \ref{sec-Discrete} to estimate $\theta$ in the three models studied in this work.

\section{Simulation Study}\label{sec-simulation}
This section is devoted to presenting the results of a simulation study. We apply the methods developed in \ref{sec-MLE} to simulated data. The purpose of this is to calibrate the estimation.
\subsection{Complete observation}\label{simulation_cont}
In this subsection, we consider the case where the data consists of path observations at a ``continuous" (complete data) time for every of the three diffusion processes considered in this work. Using Milstein Scheme, we generate paths of the diffusion process with given parameters; afterward, based on these paths, we obtain the corresponding estimators of the parameters. In particular, we show numerically the consistency of our estimators.
\subsubsection{Gompertz model}

To illustrate the consistency of the Gompertz model estimators,  we first generate paths with different time horizons. Indeed, we generate a path in the interval of time $[0,10]$ with a discretization of $\Delta_n=0.001$ and $n=10,000$. We calculated the estimators for every $100$ observations. The path was generated with $x_0=0.001$, $b=0.6$ and $\sigma=0.1$. In Figure \ref{fig:Consist} are plotted  the estimators, which were obtained using time horizons $[0,t_k]$ for $t_k=\Delta_nk$, $k=100,200,\ldots,10,000$.

\begin{figure}[H]
\centering

{\includegraphics[width=1.0\textwidth]{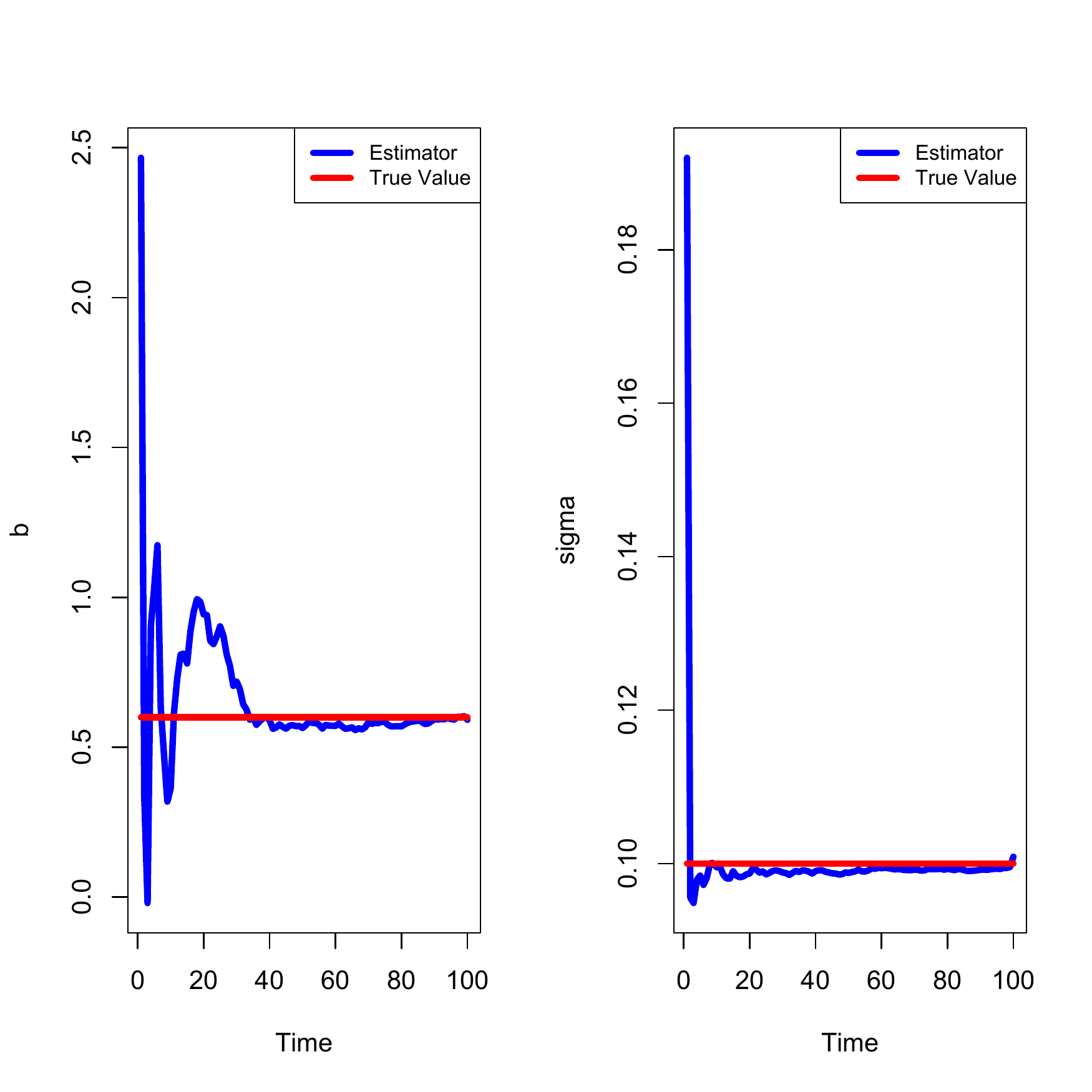}}
\caption{Consistency of estimators $b$ and $\sigma$ for the Gompertz model.}
\label{fig:Consist}
\end{figure}
On the other hand, we simulated the data set of $1,000$ paths with the same parameters $x_0,b,\sigma$, $\Delta_n$, and $n$ in the time interval $[0,10]$. The average and quantiles (95\%) of the parameter estimator are presented in Table \ref{table_gom_cont}.

\begin{table}[H]
\begin{center}
\begin{minipage}{178pt}
\caption{Average (estimator) and quantiles ($95\%$) of parameter estimates for Gompertz model obtained from 1,000 simulated datasets and 10,000 length of each path  in the interval time $[0,10]$.}
\label{table_gom_cont}
\end{minipage}

\hspace*{-2.5cm}
\begin{minipage}{178pt}
\begin{tabular}{c| c| c | c}
\textbf{Parameter} &\textbf{Real value} &\textbf{Estimator} & \textbf{Quantile 95\%}\\
\hline
\hline
$b$ & 0.6 & 0.59874   & (0.58007,0.61746) \\
\hline
$\sigma$ & 0.1 & 0.09989 & (0.09857,0.10126) \\
\hline
\end{tabular}

\end{minipage}
\end{center}
\end{table}

\subsubsection{Von Bertalanffy model}
In Figure \ref{fig:Consist_von} we show the asymptotic consistency of MLEs for Von Bertalanffy SDE. To create this  figure, we generate paths with different time horizons $[0,t_k]$ ($k=1,\ldots,100$) with a discretization of $\Delta_n=0.001$ and $n=10,000$. We calculated the estimators for every $100$ observations. The path was generated with $x_0=0.001$, $\kappa=0.6$ and $\sigma=0.1$.
\begin{figure}[H]
\centering

{\includegraphics[width=1.0\textwidth]{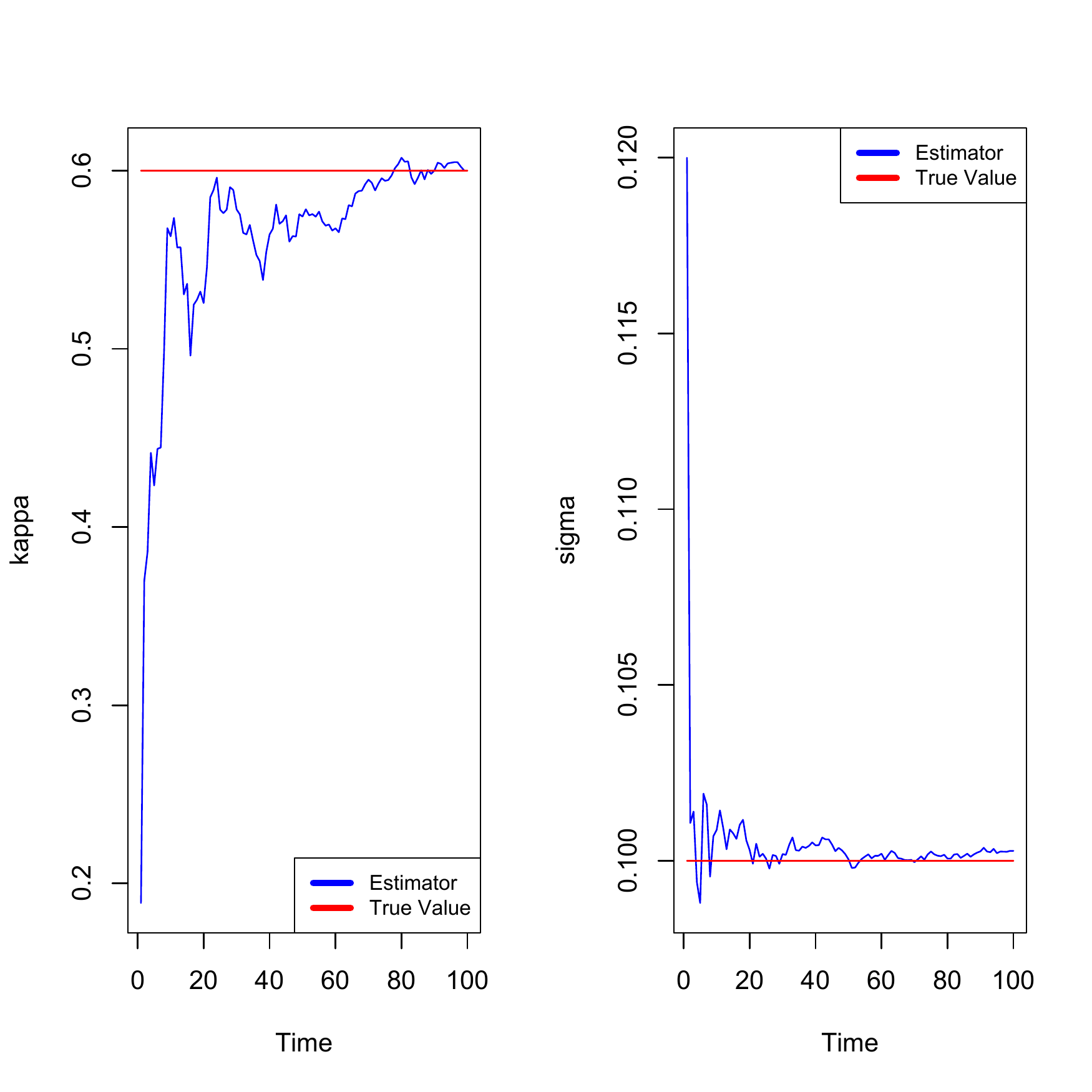}}
\caption{Consistency of estimators $b$ and $\sigma$ for the Von Bertalanffy model.}
\label{fig:Consist_von}
\end{figure}
Table \ref{table_von_cont} reports the average and quantiles (95\%) of parameter estimators obtained from a simulated dataset of 1,000 trajectories with parameters $x_0=0.001,\kappa=0.6,\sigma=0.1$, and $\Delta=0.001$ in the time interval $[0,100]$.

\begin{table}[H]
\begin{center}
\begin{minipage}{174pt}
\caption{Average (estimator) and quantiles ($95\%$) of parameter estimates for the Von Bertalanffy model, obtained from 1,000 simulated datasets and 10,000 lengths of each path  in the interval time $[0,100]$.}
\label{table_von_cont}
\end{minipage}

\hspace*{-2.5cm}\begin{minipage}{174pt}
\begin{tabular}{c| c| c | c}
\textbf{Parameter} &\textbf{Real value} &\textbf{Estimator} & \textbf{Quantile 95\%}\\
\hline
\hline
$r$ & 0.6 & 0.60086  & (0.54117,0.64894) \\
\hline
$\sigma$ & 0.1 & 0.10017 & (0.098288,  0.10213) \\
\hline
\end{tabular}

\end{minipage}
\end{center}
\end{table}

\subsubsection{Logistic model }
For this model, the chosen value for the drift parameter is $r=0.6$. We generate the paths of the SDE \eqref{sto-log} using the same methods that were described for  the other two models. We illustrated the consistency of $r$ and $\sigma$ by using an arbitrary path, the results are plotted in Figure \ref{fig:Consist_log}. We can see that the estimators stabilize close to the parameter when the observation time increases.  Based on this observation, the time horizon is chosen to simulate several paths. Table \ref{table_log_cont} reports the average and quantiles (95\%) of  the corresponding estimators. we can see that there is no bias in the estimates.
\begin{figure}[H]
\centering

{\includegraphics[width=1.0\textwidth]{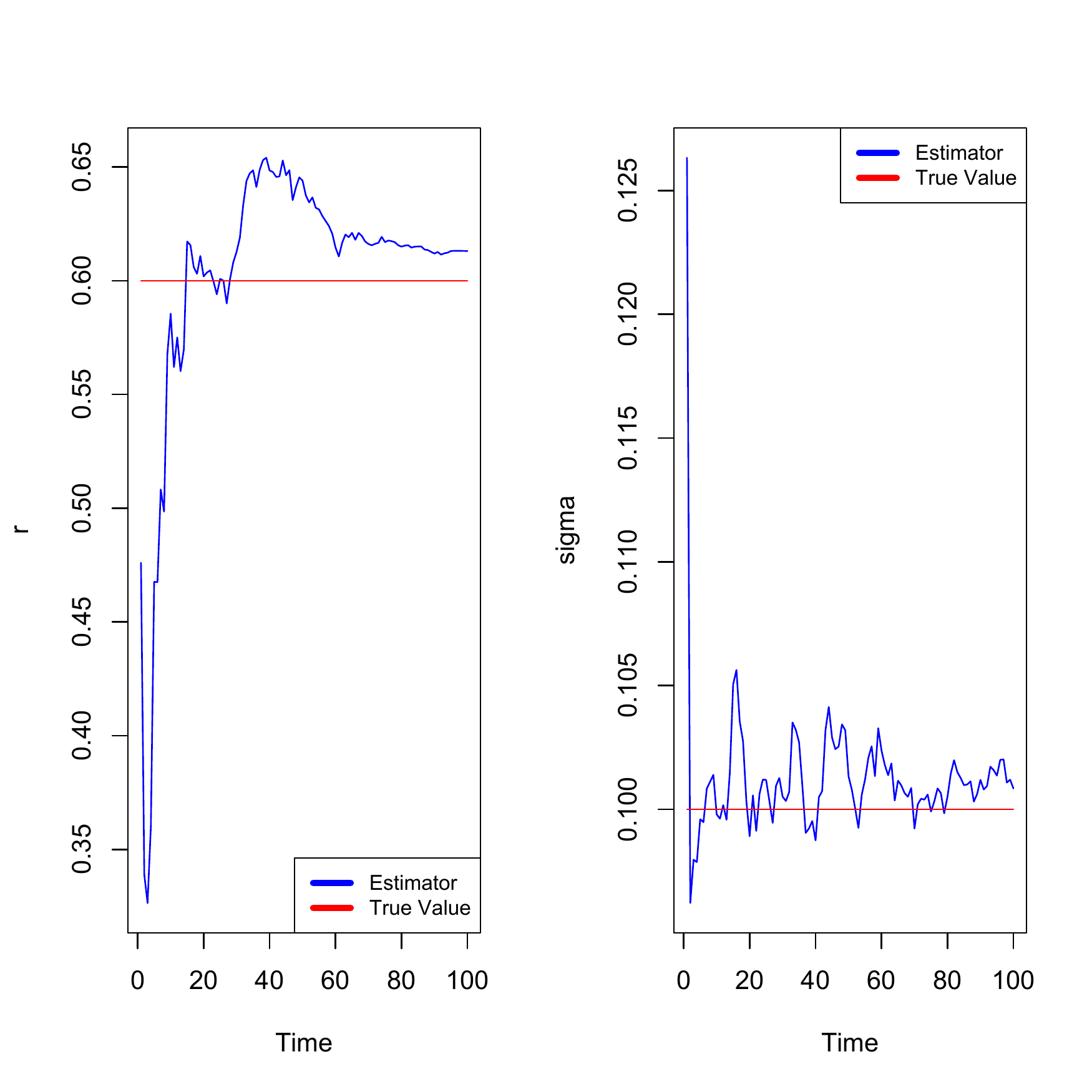}}
\caption{Consistency of estimators $r$ and $\sigma$ for the Logistic model.}
\label{fig:Consist_log}
\end{figure}

\begin{table}[H]
\begin{center}
\begin{minipage}{174pt}
  \caption{Average (estimator) and quantiles ($95\%$) of parameter estimates for the Logistic model, obtained from 1,000 simulated datasets and 10,000 lengths of each path  in the interval time $[0,100]$.}
  \label{table_log_cont}
 \end{minipage}

\hspace*{-2.5cm}\begin{minipage}{174pt}
\begin{tabular}{c| c| c | c}

\textbf{Parameter} &\textbf{Real value} &\textbf{Estimator} & \textbf{Quantile 95\%}\\
\hline
\hline
$r$ & 0.6 & 0.59405  & (0.54464,0.64755) \\
\hline
$\sigma$ & 0.1 & 0.10006 & (0.98728,0.10151) \\
\hline
\end{tabular}
\end{minipage}
\end{center}
\end{table}
\subsection{Discrete observations}\label{simulation_disc}
In this subsection, we present the validation of our methods assuming discrete observation for the three models. For this study,  $1,000$ datasets of each model were simulated. Each dataset was obtained by simulating a sample path of length $10,001$ in the interval time $[0,100]$ with an initial value $x_0=0.001$. We suppose that we have observed only $101$ points of the path at times $0=t_0,1=t_1,\ldots,t_
{1000}=100$ ($n=100$). We assume that $\Delta_i=\frac{100}{n}=1$ for all $i=1,\ldots,n$, and $L_i=10,000$ for $i=1,\ldots,n$ for each path. 

\subsubsection{Gompertz model}
In this subsection, we present the result of a simulation study for the Gompertz model that satisfies the SDE \eqref{SGE}. The given parameter values were $b=0.6$ and $\sigma=0.1$. We ran Algorithm \ref{EM-dis-gom} with the initial values obtained using the expressions \eqref{b}, and \eqref{sigma_gom} using the incomplete observation, then we have that $b_0=0.6934$, and $\sigma_0=0.2123$. Figure \ref{fig:EM-gomp} plots  the estimators of 1,000 iterations of Algorithm \ref{EM-dis-gom} for a path of the solution of SDE \eqref{SGE}.

\begin{figure}[H]
\centering
{\includegraphics[width=1.0\textwidth]{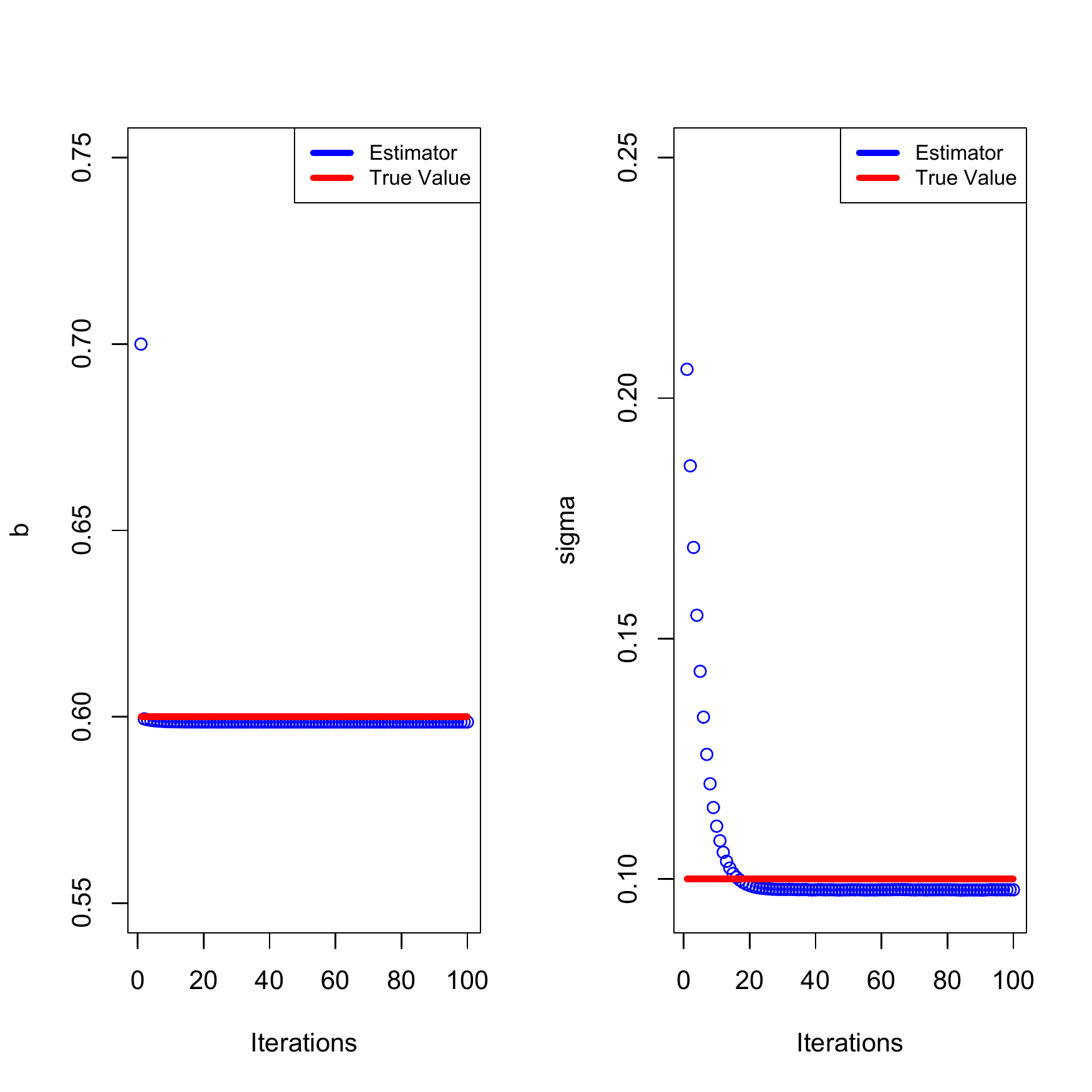}}
\caption{Iterations of EM algorithm to estimate the parameters $b$ and $\sigma$ for the stochastic Gompertz model.}
\label{fig:EM-gomp}
\end{figure}
The average and quantiles ($95\%$) of parameter estimators obtained from the sample at the $100ht$ iteration are presented in Table \ref{table_gom_dis}.

\begin{table}[H]
\begin{center}
\begin{minipage}{174pt}
  \caption{Average (estimator) and quantiles ($95\%$) of parameters for the Gompertz model.}
\label{table_gom_dis}
\end{minipage}

\hspace*{-2.5cm}\begin{minipage}{174pt}
\begin{tabular}{c| c| c | c}
\textbf{Parameter} &\textbf{Real value} &\textbf{Estimator} & \textbf{Quantile 95\%}\\
\hline
\hline
$b$ & 0.6 & 0.59943  & (0.59652,0.60286) \\
\hline
$\sigma$ & 0.1 & 0.10059 & (0.09958,0.10103) \\
\hline
\end{tabular}
\end{minipage}
\end{center}
\end{table}

\subsubsection{Von Bertalanffy model}
We now present the corresponding results for the Von Bertalanffy model. For this case, the parameter values were fixed to $\kappa=0.6$ and $\sigma=0.1$. Figure \ref{fig:dis_von} shows $100$ iterations of Algorithm \ref{EM-dis-von}. The initial values $\kappa_0=0.2083,\sigma_0=0.1461$ were obtained with the expression \eqref{sigma_van}, and \eqref{kappa_van}. 

\begin{figure}[H]
\centering

{\includegraphics[width=1.0\textwidth]{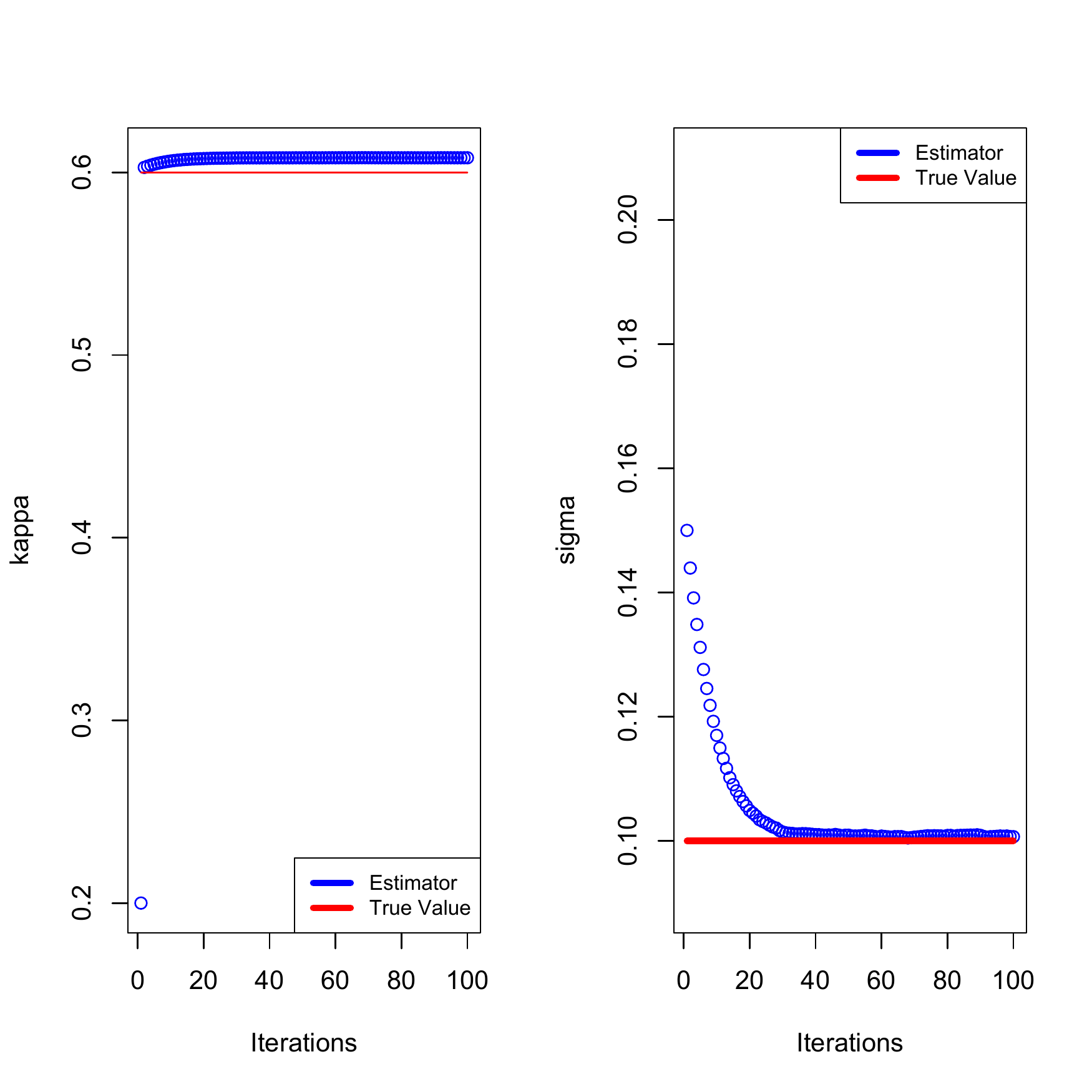}}
\caption{Iterations of EM algorithm \ref{EM-dis-von} to estimate the parameters $\kappa$ and $\sigma$ for the stochastic Von  Bertalanffy model. }
\label{fig:dis_von}
\end{figure}
Table \ref{table_von_dis} reports the average and the quantiles ($95\%$) of the last iterations of the datasets of $1,000$ paths observed at discrete time.

\begin{table}[H]
\begin{center}
\begin{minipage}{174pt}
  \caption{Average (estimator) and quantiles ($95\%$) of parameters $\kappa$ and $\sigma$ in the Von Bertalanffy model.}
\label{table_von_dis}
\end{minipage}

\hspace*{-2.5cm}\begin{minipage}{174pt}
\begin{tabular}{c| c| c | c}
\textbf{Parameter} &\textbf{Real value} &\textbf{Estimator} & \textbf{Quantile 95\%}\\
\hline
\hline
$\kappa$ & 0.6 &  0.59511 &        (0.57946,0.60992)  \\ 
\hline
$\sigma$ & 0.1 & 0.10381    &   (0.09938,0.10761)\\
\hline
\end{tabular}
\end{minipage}
\end{center}
\end{table}

\subsubsection{Logistic model}
This subsection is devoted to showing the results of a simulation study for the Logistic stochastic model when we have observations at discrete times. We present results with parameter values of  $r=0.6$ and $\sigma=0.1$. First, we illustrate Algorithm \ref{EM1} in  Figure \ref{fig:log_dis} with $100$ iterations with an arbitrary path, and arbitrary initial values $r_0=0.5,\sigma_0=0.2$. We can see that Algorithm \ref{EM1} converges very quickly and there is no bias in the estimates according to the average and the quantiles ($95\%$)  that are presented in Table \ref{table_log_dis}.

\begin{figure}[H]
\centering

{\includegraphics[width=1.0\textwidth]{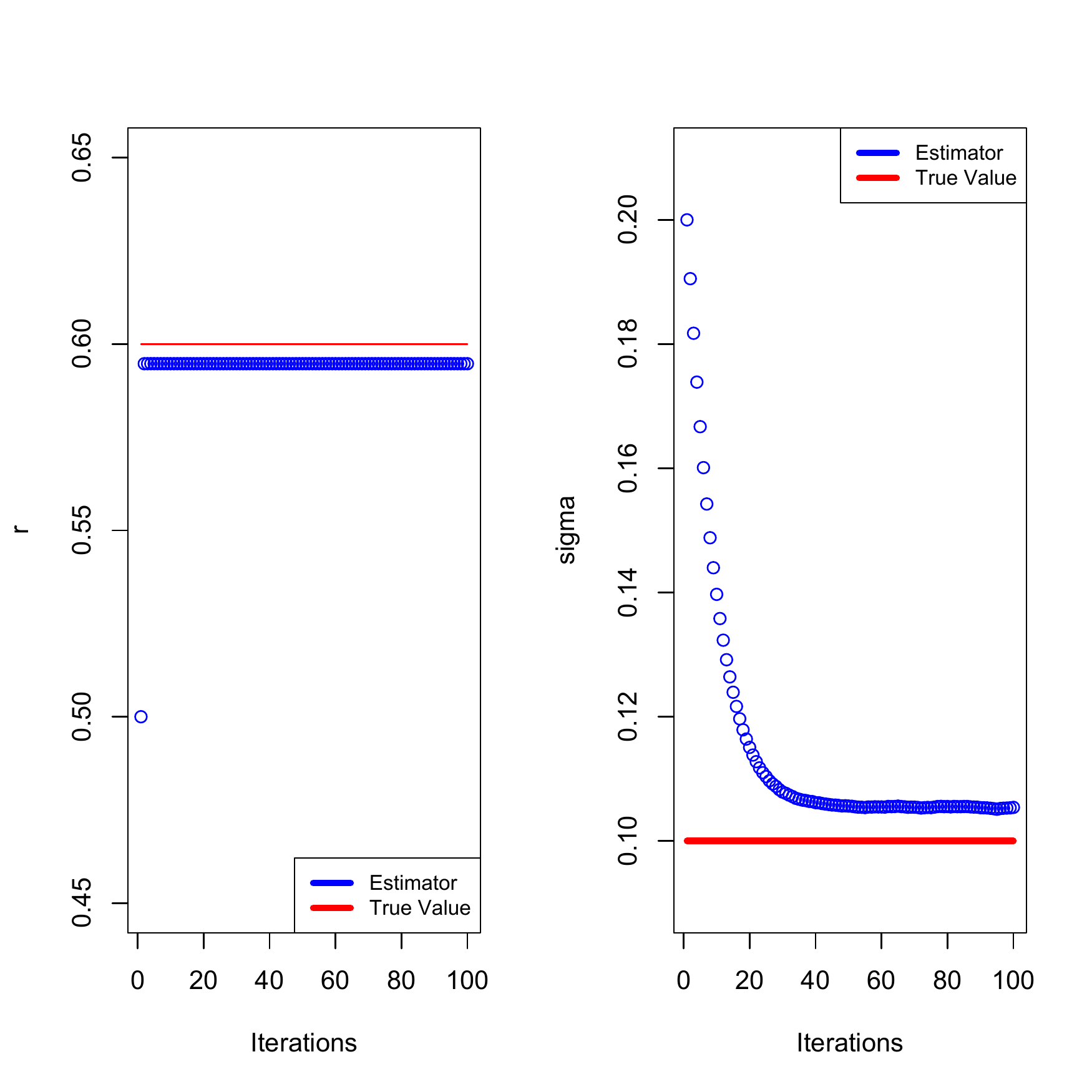}}
\caption{Iterations of EM algorithm \ref{EM1} to estimate the parameters $r$ and $\sigma$ for the stochastic Logistic model}
\label{fig:log_dis}
\end{figure}

\begin{table}[H]
\begin{center}
\begin{minipage}{174pt}
  \caption{Average (estimator) and quantiles ($95\%$) of parameter estimates  for the Logistic model, obtained from a sample of size $1,000$ of the $100ht$ iteration of Algorithm \ref{EM1}}
\label{table_log_dis}
\end{minipage}

\hspace*{-2.5cm}\begin{minipage}{174pt}
\begin{tabular}{c| c| c | c}
\textbf{Parameter} &\textbf{Real value} &\textbf{Estimator} & \textbf{Quantile 95\%}\\
\hline
\hline
$r$ & 0.6 &  0.58324 &        (0.52946,0.61549)  \\ 
\hline
$\sigma$ & 0.1 & 0.10923    &   (0.09323,0.11314)\\
\hline
\end{tabular}
\end{minipage}
\end{center}
\end{table}

\subsection{Validation of the method assuming one record }\label{simulation_one}

In this section we present a simulation study for the extreme case, in which we get only one observation of different paths; here, however, we assume we have observed several paths from the same SDE. We assumed they are observed  at discrete times.

\subsubsection{Gompertz model}

We used Algorithm \ref{OR} to generate a path of a process solution of the equation \eqref{SGE}. Algorithm was run for $M=100$, $n=10,000$ and $\Delta_i=0.001$ for all $i=1,\ldots,n$. We fixed the parameters, to generate each path, as $b=0.6$ and $\sigma=0.1$ with initial distribution $Beta(1,100)$.  We suppose that we have only $1001$ observations at times $0=t_0,1=t_1,\ldots,t_
{1000}=1$ ($k=1000$). Then $\Delta_k=0.01$ for each path. Thus, using Algorithm \ref{OR} we have a trajectory observed at discrete time and we can use Algorithm \ref{EM-dis-gom} to find the corresponding MLEs. We run Algorithm \ref{EM-dis-gom} with 50 iterations and initial values $b_0=0.5$ and $\sigma_0=0.2$. Figure \ref{fig:gom_one} shows the iterations of the Algorithm \ref{EM-dis-gom}.

\begin{figure}[H]
\centering

{\includegraphics[width=1.0\textwidth]{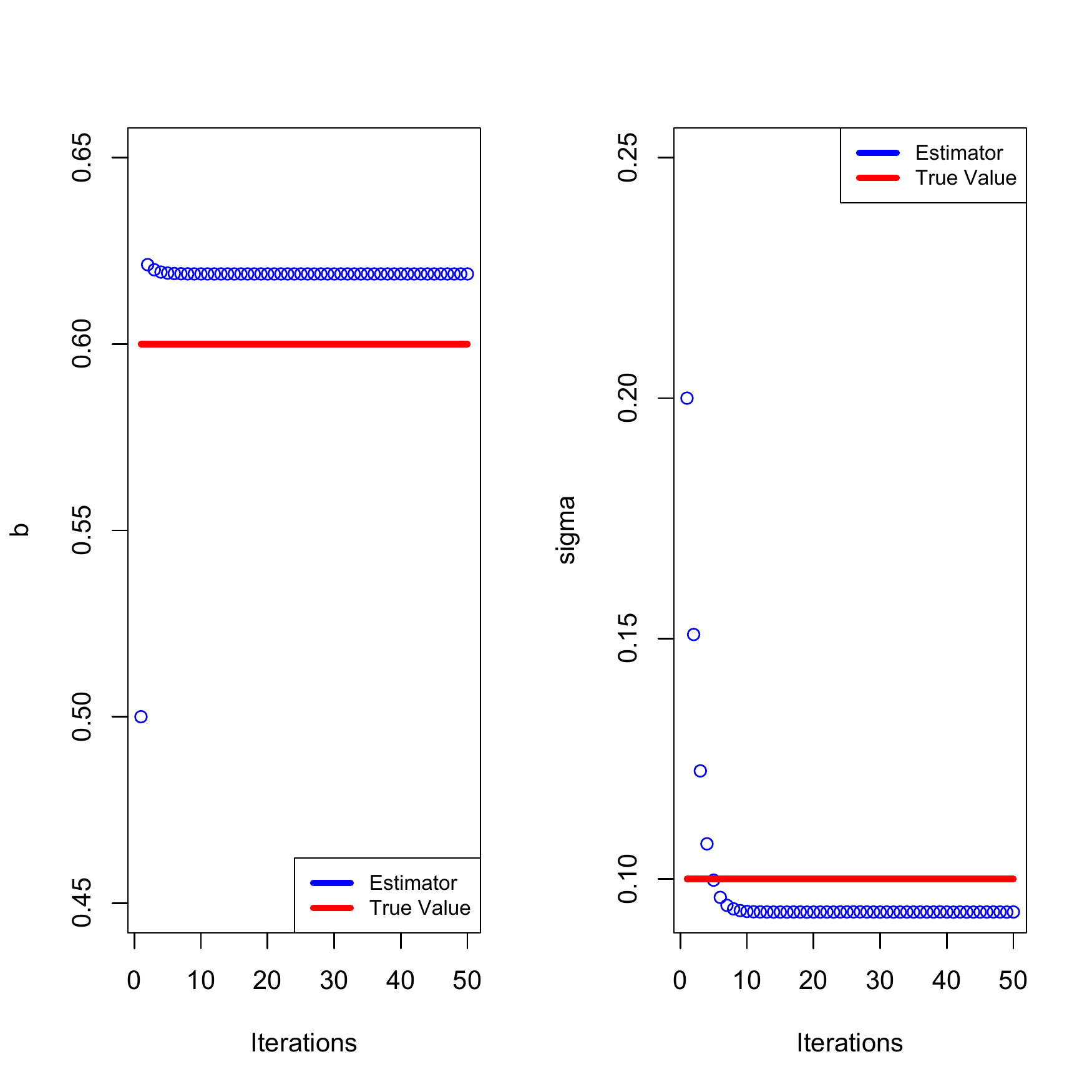}}
\caption{ Iterations of EM algorithm \ref{EM-dis-gom} to estimate the parameters $b$ and $\sigma$ for the stochastic Gompertz model.}
\label{fig:gom_one}
\end{figure}

\subsubsection{The Von Bertalanffy model}

We now apply Algorithm \ref{OR} to generate a path observed at discrete time in Von Bertalanffy model where the real values of parameters are $\kappa=0.6$ and $\sigma=0.1$  and the initial value follows a $Beta(1,100)$. We run Algorithm \ref{OR} with $M=100$, $n=10,000$ and $\Delta_i=0.001$  where we have only $1001$ observations at times $0=t_0,1=t_1,\ldots,t_
{1000}=1$ ($k=1000$). Using the path generated by Algorithm \ref{OR} we apply Algorithm \ref{EM-dis-von} to find the corresponding MLEs. Algorithm \ref{EM-dis-von}  was run with 50 iterations and initial values $\kappa_0=0.4$ and $\sigma_0=0.25$. In Figure \ref{fig:von_one} we can observe the evolution of Algorithm \ref{EM-dis-von}.

\begin{figure}[H]
\centering

{\includegraphics[width=1.0\textwidth]{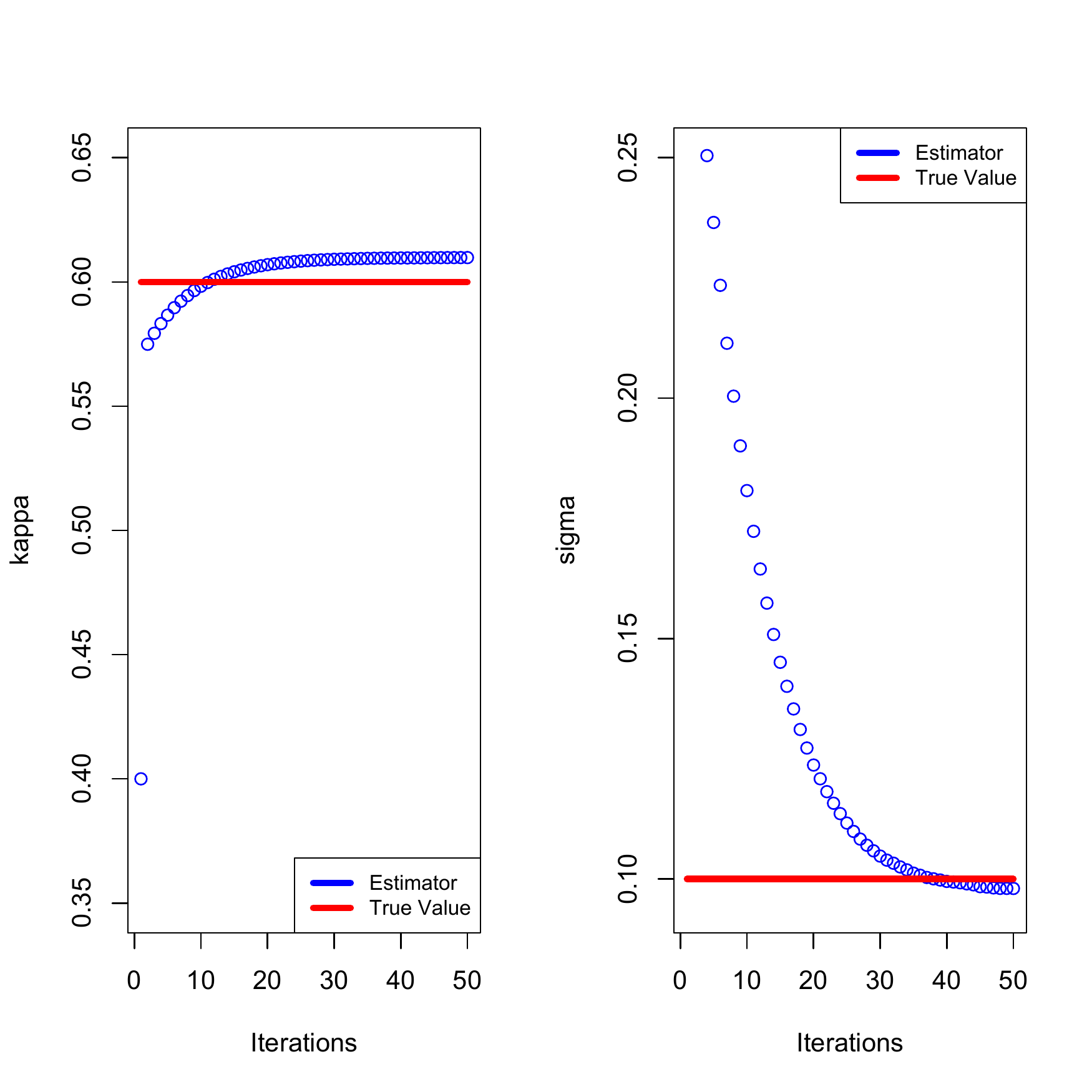}}
\caption{Iterations of EM algorithm \ref{EM-dis-von} to estimate the parameters $\kappa$ and $\sigma$ for the stochastic Von Bertalanffy model.}
\label{fig:von_one}
\end{figure}

\subsubsection{Logistic model}

Finally, we make the same simulation study for the Logistic model. The corresponding parameters  to generate data are $r=0.6$ and $\sigma=0.1$. The other parameters of  Algorithm \ref{OR} are the same that we used in the two sections previous. Algorithm \ref{EM1}  was run with 50 iterations and initial values $r_0=0.8$ and $\sigma_0=0.3$. Figure \ref{fig:log_one} shows the obtained results.

\begin{figure}[H]
\centering

{\includegraphics[width=1.0\textwidth]{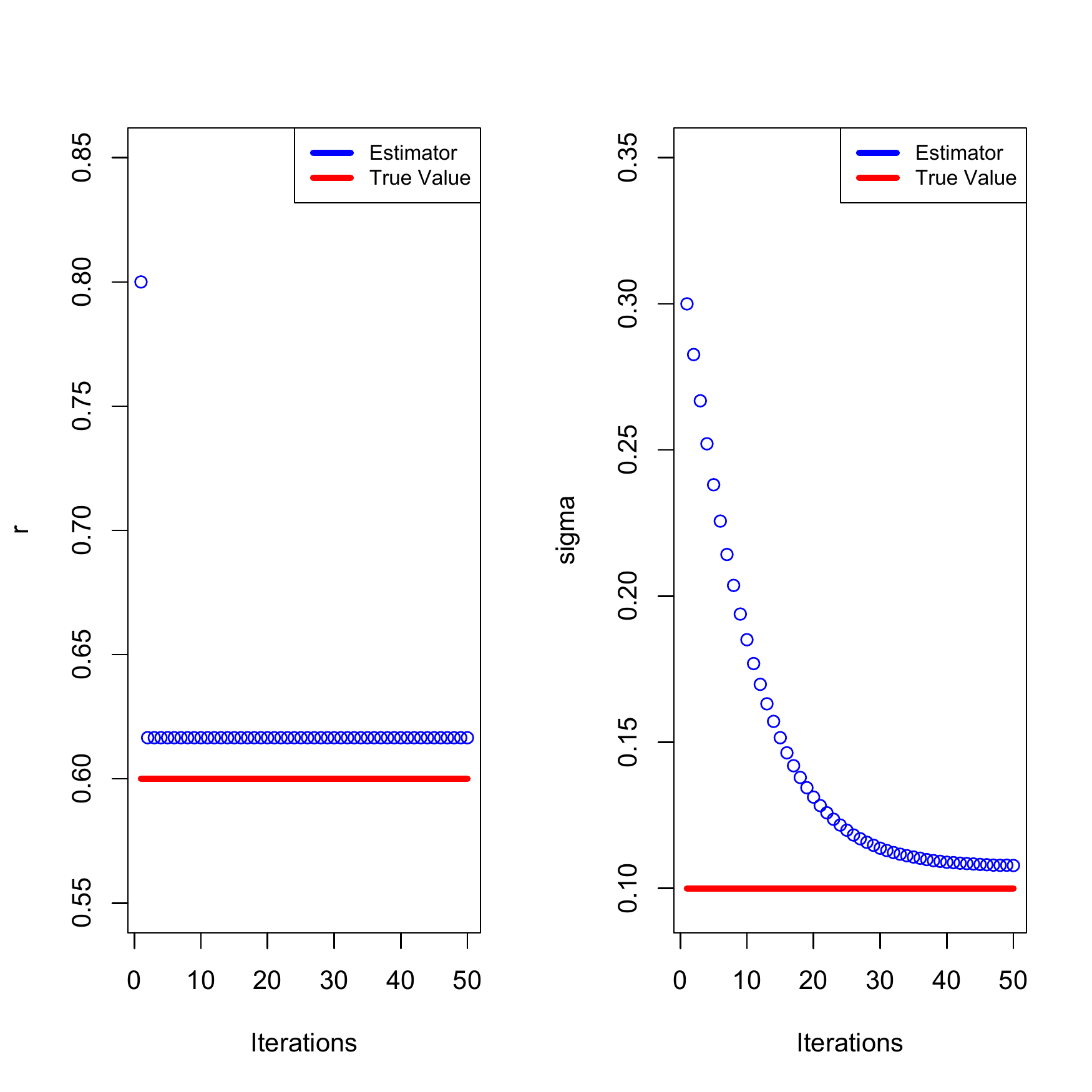}}
\caption{Iterations of EM algorithm \ref{EM1} to estimate the parameters $r$ and $\sigma$ for the stochastic Logistic model. }
\label{fig:log_one}
\end{figure}

\section{Model selection via AIC}
\label{sect_model_sel}

\subsection{Estimators for each path and model.}

Akaike's information criterion, referred to as AIC, is a method for evaluating and comparing statistical models developed by the Japanese mathematician Hirotsugu Akaike (see \cite{aka}). It provides a measure of the quality of the estimate of a statistical model taking into account both the goodness of fit and the complexity of the model. \cite{bu-an}. The well-known Akaike information criterion is given by
\begin{align}
\mbox{AIC} = -2\log L(\hat{\theta}\mid \mbox{data}) + 2\kappa_\theta ,
\end{align}
where $L(\hat\theta \mid \mbox{data}) $ is the likelihood, $\hat\theta$ is an estimator for the parameter (or parameters) and $\kappa_\theta$ is the total number of parameters in the model. Usually, $\hat\theta $  is taken as the maximum likelihood estimator  $\hat\theta_{MLE}$.

Consider the two stochastic differential equations:

\begin{align}\label{gen_SDE_1}
d X_t &= \alpha F(X_t) dt + \sigma G(X_t) dB_t,\\
X_0&=x_0,\nonumber
\end{align}

and 

\begin{align}\label{gen_SDE_2}
d X_t &= \sigma G(X_t) dB_t,\\
X_0&=x_0.\nonumber
\end{align}
Denote by $\Pb_\alpha$ and $\Pb$ the probability measures induced by each solution of the equations, respectively. 

Thus, as was mentioned before, $\Pb_\alpha$ and $\Pb$ are equivalent (see \cite{iacus}) and the corresponding Radon-Nikodym derivative, which means the Likelihood, is given by the Girsanov theorem

\begin{align}
\frac{d\Pb_\alpha}{d\Pb}(X)=\exp\left[\int_0^T  -\alpha \frac{F(X_t)}{\sigma^2 G^2(X_t)} dX_t - \frac{1}{2} \int_0^T  \alpha^2 \frac{F^2(X_t)}{\sigma^2 G^2(X_t)} dt \right].
\end{align}

Then, the log-likelihood is defined as
\begin{align}\label{Likelihood_SDE_Gen}
\log L(\alpha)= \int_0^T  -\alpha \frac{F(X_t)}{\sigma^2 G^2(X_t)} dX_t - \frac{1}{2} \int_0^T  \alpha^2 \frac{F^2(X_t)}{\sigma^2 G^2(X_t)} dt. 
\end{align}

Thus, the AIC for the stochastic model \eqref{gen_SDE_1} is
\begin{align}\label{AIC_SDE_Gen}
\mbox{AIC}= - 2 \log L(\hat\alpha_{ML}) + 2k,
\end{align}
with $k$ being the number of estimated parameters in the model.

\begin{remark}

To obtain the AIC for the stochastic Gompertz model we apply  \eqref{Likelihood_SDE_Gen} with $F(x)=x\log(x)$ and $G(x)=x$. \\

The AIC for the stochastic Von Bertalanffy model we apply  \eqref{Likelihood_SDE_Gen} to the functions $F(x)=G(x)=L_\infty -x$.\\

Finally, the AIC for the stochastic Logistic model we use  \eqref{Likelihood_SDE_Gen} with the functions $F(x)=x(1-x)$ and $G(x)=x$.
\end{remark}

\subsection{A numerical example}
This section is devoted to studying a numerical example of the AIC implementation for the three SDEs . We proceed as follows. First, we generate a sample path for each SDE studied in this work. Afterward, we use our methods to estimate the corresponding parameter for each SDE. At this point, we assume we do not know what is the "true" model, then we fit every sample path for the three SDEs. This implies that we have for every sample path three possible models and to choose the best model we apply the AIC criteria. To do that, we calculate the AIC for each case and verify that the "true" model is the one that minimizes the AIC.\\

We generate the three paths using the same initial value $x_0=0.01$, for the Gompertz model the parameter $b=0.6$, for the Von Bertalanffy  $\kappa=0.6$, and $r=0.6$ in Logistic case. The value of the parameter $\sigma=0.1$ for all models and we generate the paths in the time interval $[0,10]$ with a discretization of $\Delta_n=10/n$ where $n=10,000$. Figure \ref{fig:paths} shows paths for all models using the corresponding parameters.

\begin{figure}[H]
\centering

{\includegraphics[width=1.0\textwidth]{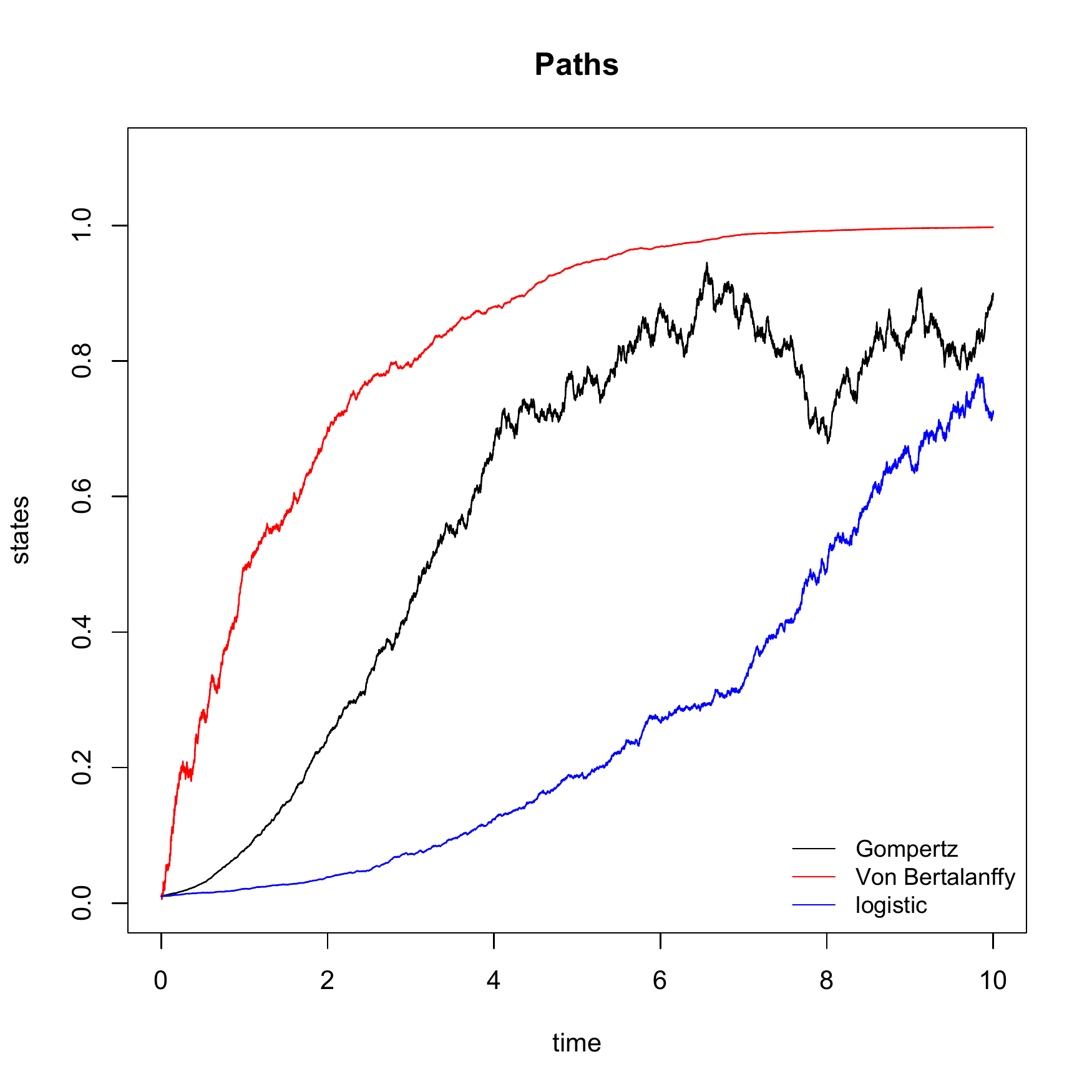}}
\caption{Paths}
\label{fig:paths}
\end{figure}

Table \ref{Estimations} reports the value of the parameters of every model with the description given above. We use these parameters to calculate the AIC. We observe that we are assuming that the log-likelihood is a function only of the drift parameters.

\begin{table}[H]
\begin{center}
\begin{minipage}{194pt}
 \caption{Estimators fitted for each model.}
\label{Estimations}
\end{minipage}

\hspace*{-7cm}\begin{minipage}{174pt}
\begin{tabular}{|c| c| c | c | c | c}
\hline
\textbf{Parameter} & \textbf{True Model} &  \textbf{Fitted model} &  \textbf{Real value} &\textbf{Estimator} \\
\hline
\hline
$b$ & Gompertz  & Gompertz  & 0.6 & 0.60443  \\
$\sigma$ & Gompertz  & Gompertz  & 0.1 & 0.09982\\
$b$ &  Von Bertalanffy & Gompertz  & 0.6 & 3.91481  \\
$\sigma$ & Von Bertalanffy  & Gompertz  & 0.1 & 0.47852\\
$b$ &  Logistic & Gompertz  & 0.6 & 0.14383  \\
$\sigma$ & Logistic  & Gompertz  & 0.1 & 0.09904\\
\hline
$\kappa$ & Gompertz  & Von Bertalanffy  & 0.6 & 0.22904 \\
$\sigma$ & Gompertz  & Von Bertalanffy  & 0.1 & 0.13125\\
$\kappa$ &  Von Bertalanffy & Von Bertalanffy   & 0.6 &  0.59583  \\
$\sigma$ & Von Bertalanffy  & Von Bertalanffy   & 0.1 & 0.105623\\
$\kappa$ &  Logistic & Von Bertalanffy & 0.6 &  0.12857 \\
$\sigma$ & Logistic  & Von Bertalanffy  & 0.1 & 0.04493\\
\hline
$r$ & Gompertz  & Logistic  & 0.6 & 1.41950 \\
$\sigma$ & Gompertz  & Logistic  & 0.1 & 0.100101\\
$r$ &  Von Bertalanffy & Logistic  & 0.6 &  4.39651  \\
$\sigma$ & Von Bertalanffy  & Logistic   & 0.1 & 0.03536\\
$r$ &  Logistic & Logistic & 0.6 &  0.590984\\
$\sigma$ & Logistic  & Logistic & 0.1 & 0.10030\\
\hline
\end{tabular}
\end{minipage}
\end{center}
\end{table}

Table \ref{AIC} reports the AIC given a true model and fitting the three models to each set of the simulated data. To illustrate the table if we take the Gompertz SDE as the true model, then the AIC when it is assumed the Gompertz model is -919.09, for the Von Bertalanffy is 301.39, and for the Logistic is -283.42. Thus, from the results contained in this table, we conclude that in the three cases, the AIC criterion allows us to choose the ``true" model.\\

\begin{table}[H]
\begin{center}
\begin{minipage}{194pt}
  \caption{AIC for each model and adjusted model.}
\label{AIC}
\end{minipage}

\hspace*{-7cm}\begin{minipage}{174pt}
\begin{tabular}{|c|c| c| c |}
\hline
\textbf{Fitted Model - True Model}  &\textbf{Gompertz}  &  \textbf{Von Bertalanffy} & \textbf{Logistic} \\
\hline
\textbf{Gompertz}  & \textbf{-919.09} & -30.44 & 1534.14 \\
\textbf{Von Bertalanffy} &301.39    & \textbf{-318.18} &  38389.18\\   
\textbf{Logistic}& -283.42 &-81.88 & \textbf{628.95}\\
\hline

\end{tabular}

\end{minipage}
\end{center}
\end{table}

Finally, we replicate the previous experiment by simulating $10,000$  paths for each process within the time interval $[0,10]$, and with a discretization of $\Delta_n=10/n$, where $n=10,000$ (continuous case). We have used the same initial condition for each simulation but different for each SDE. Let us denote by $pc$ the probability of choosing the correct model; then to estimate the probability $pc$, we divide the number of times the correct model was chosen ($nc$) by the total number of experiments conducted ($ne$),  that is $pc=nc/ne$. Table \ref{MCAIC} reports the results of the simulation. From the data presented in Table \ref{MCAIC}, it becomes evident that the selection of the model using the AIC is exceptionally effective.
\begin{table}[H]
\begin{center}
\begin{minipage}{194pt}
  \caption{Probability of choosing the correct model.}
\label{MCAIC}
\end{minipage}

\hspace*{-8cm}\begin{minipage}{174pt}
\begin{tabular}{|c|c|  c|c|}
\hline
\textbf{Model}  
&  \textbf{Drift parameter} & \textbf{Diffusion parameter} & \textbf{$pc$} \\
\hline
\textbf{Gompertz}  &
 0.6 & 0.006 & 0.9710\\
\textbf{Von Bertalanffy}  & 0.6 &  0.06 & 0.9689\\   
\textbf{Logistic}& 0.6& 0.05& 0.9599 \\
\hline

\end{tabular}

\end{minipage}
\end{center}
\end{table}

\section{Concluding remarks} \label{sec-Conclusions}

In this paper, we have presented a method to fit SDEs to different types of datasets as follows. To illustrate the model, we consider as examples three stochastic models for biological growth. Each model is given by different stochastic differential equations which are a stochastic version of classical deterministic differential equations, which have been applied to several fields of science, and therefore they are very important. We have considered SDEs driven by an affine noise.\\ 

 We assumed that fitting the stochastic model is equivalent to adjusting only a fixed number of parameters in each SDE. Furthermore, we considered that these parameters are constants but unknown and we estimated them. We have shown a method to estimate these parameters for different types of datasets. Indeed, the dataset we successfully managed could be a path of the SDE that is either discrete or continuous. Moreover, the dataset could have only one observation for each path, subject to the condition that we have a sufficient number of observed paths. This permits us to rebuild (continuous or discrete) paths to estimate the parameters. \\

Finally, in the method presented here, we have used the classical Akaike information criterion (AIC) to select the best model. We have used Girsanov's Theorem to define the log-likelihood and thus the AIC. We have run simulations to validate the estimation procedure and the selection of the best model.  For the simulated data, we have found that the AIC provides, as in the deterministic case, a good tool for selecting models. 

\section{Acknowledgements}
Baltazar-Larios F. has been supported by UNAM-DGAPA-PASPA.

\section{Declarations}
\textbf{Conflict of interest} The authors declare that they have no competing interests.



\begin{thebibliography}{}



\bibitem[Ando, 2010]{ando-10}
Ando, T. (2010).
\newblock {\em Bayesian model selection and statistical modeling}.
\newblock North Carolina State University, North Carolina State University.

\bibitem[Bladt and M., 2014]{bla-sor-14}
Bladt, M. and  S\o rensen, M. (2014).
\newblock Simple simulation of diffusion bridges with application to likelihood
  inference for diffusions.
\newblock {\em Bernoulli}, 20(2):645--675.

\bibitem[Akaike, H., 1974]{aka}
Akaike, H. (1974).
\newblock A new look at the statistical model identification. 
\newblock {\em IEEE transactions on automatic control.}, 19(6), 716-723.

\bibitem[Braun et~al., 1983]{br-et-al-83}
Braun, M., Coleman, C.~S., Drew, D.~A., and Lucas, W.~F. (1983).
\newblock {\em Differential equation models}, volume~1.
\newblock Springer-Verlag, New York.

\bibitem[Braun and Golubitsky, 1992]{br-go}
Braun, M. and Golubitsky, M. (1992).
\newblock {\em Differential equations and their applications}, volume~1.
\newblock Springer-Verlag, New York, 4 edition.

\bibitem[Burnham and Anderson, 2002]{bu-an}
Burnham, K.~P. and Anderson, D.~R. (2002).
\newblock {\em Model selection and multimodel inference}.
\newblock A practical information-theoretic approach. Springer-Verlag, New
  York.

\bibitem[Celeux and Diebolt, 1986]{ce-di-86}
Celeux, G. and Diebolt, J. (1986).
\newblock The sem algorithm: A probabilistic teacher algorithm derived from the
  em algorithm for mixture problem.
\newblock {\em Comput. Statist}, pages 599--613.

\bibitem[Chao and Huisheng, 2016]{Chao-16}
Chao, W. and Huisheng, S. (2016).
\newblock Maximum likelihood estimation for the drift parameter in diffusion
  processes.
\newblock {\em Stochastics}, 88(5):699--710.


\bibitem[DeAngelis and Gross, 2018]{de-Gross}
DeAngelis, D.~L. and Gross, J. (2018).
\newblock {\em Individual-based models and approaches in ecology: populations,
  communities and ecosystems}.
\newblock CRC Press, United States.


\newblock {\em Scandinavian Journal of Statistics}, 40(2):322--343.

\bibitem[Delgado-Vences et-al., 2022]{de-etal-22}
Delgado-Vences, F., Baltazar-Larios, F., Ornelas-Vargas, A.,
  Morales-Boj{\'o}rquez, E., Cruz-Escalona, V.~H., and Salom{\'o}n~Aguilar, C.
  (2022).
\newblock Inference for a discretized stochastic logistic differential equation
  and its application to biological growth.
\newblock {\em Journal of Applied Statistics}.

\bibitem[Dempster et~al., 1977]{de-La-lr}
Dempster, A., Laird, N., and D.B., R. (1977).
\newblock Maximum likelihood from incomplete data via the em algorithm (with
  discussion).
\newblock {\em J. R. Stat. Soc., Ser. B Stat. Methodol}, 39:1--38.




\bibitem[Hasanoglu and Romanov, 2017]{ha-ro}
Hasanoglu, A.~H. and Romanov, V.~G. (2017).
\newblock {\em Introduction to inverse problems for differential equations}.
\newblock Springer International Publishing, New York.

\bibitem[Iacus, 2009]{iacus}
Iacus, S.~M. (2009).
\newblock {\em Simulation and inference for stochastic differential equations:
  with R examples}.
\newblock Springer Science \& Business Media, New York.

\bibitem[isa, 2006]{isakov}
 Isakov, V.  (2006).
\newblock {\em Inverse problems for partial differential equations}, volume
  127.
\newblock Springer, New York.

\bibitem[Jiang and Shi, 2005]{ji-shi}
Jiang, D. and Shi, N. (2005).
\newblock A note on nonautonomous logistic equation with random perturbation.
\newblock {\em Journal of Mathematical Analysis and Applications},
  303(1):164--172.

\bibitem[Lahouel et al, 2022]{la-etal-22} Lahouel, K., Wells, M., Lovitz, D., Rielly, V., Lew, E., and Jedynak, B. (2022). Learning
Nonparametric Ordinary differential Equations: Application to Sparse and Noisy Data. arXiv preprint
arXiv:2206.15215

\bibitem[Lillacci and Khammash, 2010]{Lilla-10}
Lillacci, G. and Khammash, M. (2010).
\newblock Parameter estimation and model selection in computational biology.
\newblock {\em PLOS Computational Biology}, 6(3):1--17.

\bibitem[Moss, 2021] {moss-etal} Moss, D. K.; Ivany, L. C.; Jones, D. S. (2021). \newblock Fossil bivalves and the sclerochronological reawakening. \newblock {\em Paleobiology}: 1–23

\bibitem[Nielsen, 2000]{ni-00}
Nielsen, S. F. . (2000).
\newblock The stochastic em algorithm: estimation and asymptotic results.
\newblock {\em Bernoulli}, 6(3):457--489.


\bibitem[Paek and Choi, 2014]{Paek-14}
Paek, J. and Choi, I. (2014).
\newblock Bayesian inference of the stochastic gompertz growth model for tumor
  growth.
\newblock {\em Communications for Statistical Applications and Methods},
  21(6):521--528.

\bibitem[Panik, 2017]{panik}
Panik, M.~J. (2017).
\newblock {\em Stochastic Differential Equations: An Introduction with
  Applications in Population Dynamics Modeling}.
\newblock John Wiley \& Sons, United Kingdom.

\bibitem[Pavliotis, 2014]{pavliotis}
Pavliotis, G.~A. (2014).
\newblock {\em Stochastic processes and applications: diffusion processes, the
  Fokker-Planck and Langevin equations}, volume~60.
\newblock Springer, New York.



\bibitem[Román-Román et~al., 2010]{ROMANROMAN201059}
Román-Román, P., Romero, D., and Torres-Ruiz, F. (2010).
\newblock A diffusion process to model generalized von bertalanffy growth
  patterns: Fitting to real data.
\newblock {\em Journal of Theoretical Biology}, 263(1):59--69.

\bibitem[Wei-Cheng, 2006]{miao}
Wei-Cheng, M. (2006).
\newblock Estimation of diffusion parameters in diffusion processes and their
  asymptotic normality.
\newblock {\em Int. J. Contemp. Math. Sciences}, 1(16):763 -- 776.

\end{thebibliography}

\end{document}